\documentstyle[aps,multicol,epsfig]{revtex}
\def\be{\begin{eqnarray}}
\def\ee{\end{eqnarray}}

\begin{document}
\title{Reconstruction of Liouvillian Superoperators}
\author{Vladim\'{\i}r Bu\v{z}ek}
\address{
 Institute of Physics, Slovak Academy of Sciences, D\'{u}bravsk\'{a}
cesta 9, 842 28 Bratislava, Slovakia
}

\date{February 16, 1998}
\maketitle
\begin{abstract}
We show how to determine (reconstruct) a master equation
governing the time evolution of an open quantum system.
 We present a general algorithm for the reconstruction of the corresponding 
Liouvillian superoperators.  Dynamics of a two-level atom 
in various environments is discussed in detail. 
\end{abstract}

\vspace{-0.2cm}
\begin{multicols}{2}

Proper description  of quantum dynamics of open systems is essential
for our understanding  of  physical processes in many areas of
physics, starting from quantum optics to quantum cosmology.
In general an open system can be represented as a system
$S$ interacting with an environment $E$
 \cite{Davies}. In this paper we consider the archetypal
{\it system $+$ environment} model which is specified as follows:
Let ${\cal H}_{_S}$ denotes a 
Hilbert space of the system $S$, and ${\cal H}_{_E}$ is the
Hilbert space associated with the environment $E$.
The Hamiltonian
$
\hat{H}_{_{SE}}= \hat{H}_{_S}\otimes \hat{1}_{_E}
+ \hat{H}_{int} + \hat{1}_{_S}\otimes \hat{H}_{_E}
$
of the composite system $S\oplus E$
acts on ${\cal H}_{_S}\otimes{\cal H}_{_E}$. It is assumed that $S\oplus E$
is a {\it closed finite-dimensional}
 system which evolves unitarily.  The density
operator $\hat{\rho}_{_{SE}}(t)$ of this composite system is governed by the
von Neumann equation with the formal solution $\hat{\rho}_{_{SE}}(t)=
\exp[-i(t-t_0)\hat{H}_{_{SE}}]\hat{\rho}_{_{SE}}(t_0)
\exp[i(t-t_0)\hat{H}_{_{SE}}]$, 
where  the initial state is 
$\hat{\rho}_{_{SE}}(t_0)=\hat{\rho}_{_S}(t_0)\otimes\hat{\rho}_{_E}(t_0)$
and $\hbar=1$.
The {\it reduced} dynamics
 of the system $S$ is then defined as
\be
\hat{\rho}_{_S}(t) := \hat{\cal T}(t,t_0) \hat{\rho}_{_S}(t_0) =
{\rm Tr}_{_E} \left[\hat{\rho}_{_{SE}}(t)\right].
\label{1}
\ee
By definition, $\hat{\cal T}(t,t_0)$ is a linear map which transforms the
input state $\hat{\rho}_{_S}(t_0)$ onto the output state $\hat{\rho}_{_S}(t)$.
In this paper we address the question {\it how to determine (reconstruct)
the master equation which governs the time evolution of the reduced
density operator $\hat{\rho}_{_S}(t)$}.
This master equation can be written in the {\it convolutionless}
 form \cite{Bona} (we omit the subscript $S$)
\be
\frac{d}{d t}\hat{\rho}(t) = \hat{\cal L}(t,t_0)\hat{\rho}(t).
\label{3}
\ee
which is possible due to the fact that in the {\it finite-dimensional}
Hilbert spaces matrix elements of density operators are analytic
functions. Consequently, $\hat{\cal T}(t,t_0)$ are non-singular
operators (except may be for a set of {\it isolated} values of $t$)
in which case the inverse operators $\hat{\cal T}(t,t_0)^{-1}$ exist
and the Liouvillian superoperator can be expressed as
\be
\hat{\cal L}(t,t_0):= \left[\frac{d}{dt}\hat{\cal T}(t,t_0)\right]
\hat{\cal T}^{-1}(t,t_0).
\label{2}
\ee
We note that $\hat{\cal T}(t,t_0)$   is uniquelly specified by 
 $\hat{H}_{_{SE}}$ and by the initial
state $\hat{\rho}_{_E}(t_0)$ of the environment.

In this paper we propose a general algorithm how to reconstruct
the Liouvillian superoperator 
 $\hat{\cal L}(t,t_0)$  from the
the knowledge of the time evolution of the system
density operator $\hat{\rho}(t)$. In fact there are two aspects of this
problem.
Firstly,
 $\hat{\rho}(t)$ can be given as a result of a sequence of
quantum-tomography measurements \cite{Leonhardt}, such that at each time
$t$ the system density operator is reconstructed from the measured
tomographic data. From these experimental
data then the Liouvillian which governs the
open system can be {\it reconstructed} (see Example {\it A}). 
Secondly, the density operator
$\hat{\rho}(t)$ is determined from the knowledge of the unitary evolution
of the composite $S\oplus E$ system [see Eq.(\ref{1})]. From this knowledge
 the master equation (\ref{3}) is determined.
In both cases the dynamics of the open system is given exclusively in terms
of the system operators. Environmental degrees of freedom are
completely eliminated from the reduced dynamics. Nevertheless,
the state of the environment may change during the time evolution
due to the interaction with the system.
That is, we do not employ the assumption
that the environment is a ``big'' reservoir which does not change under
 the action of the system (see Examples {\it B} and {\it C}).

In order to reconstruct the Liouvillian superoperator $\hat{\cal L}(t,t_0)$
we have  to determine firstly the linear map $\hat{\cal T}(t,t_0)$ given
by Eq.(\ref{1}). This part of the reconstruction can be performed with the
help of the algorithm recently proposed by
 Poyatos, Cirac and Zoller \cite{Poyatos}. This algorithm works as follows:
Let us assume that the system $S$ has been initially   prepared
in a pure state $|\Psi(t_0)\rangle=\sum_{i_1=0}^N c_{i_1}| i_1\rangle$
where $|i_1\rangle$ are basis vectors in the $(N+1)$-dimensional
Hilbert space ${\cal H}_{_S}$ of the system under consideration.
It is further assumed that the environment is initially prepared in a
state
$\hat{\rho}_{_E}(t_0)=\sum_{\alpha_1\alpha_2} d_{\alpha_1\alpha_2}
|\alpha_1\rangle_{_E}\langle\alpha_2|$,
where $|\alpha_i\rangle_{_E}$ are	basis vectors in the Hilbert space
${\cal H}_{_E}$ of the environment.

In general, the physical process $\hat{\cal T}(t_k,t_0)$ is determined by
a transformation acting on basis vectors of the system and the environment
(in what follows we omit in all expressions
the explicit reference to the initial time $t_0$)
\be
|i_1\rangle_{_S} |\alpha_1\rangle_{_E} 
\stackrel{\hat{\cal T}(t_k)}{\longrightarrow}
\sum_{j_1=0}^N\sum_{\beta_1} E_{(i_1j_1)(\alpha_1\beta_1)}(t_k)
|j_1\rangle_{_S} |\beta_1\rangle_{_E}.
\label{4}
\ee
The output density operator $\hat{\rho}(t_k)$
of the system at time $t_k$ is	obtained when the transformation
(\ref{4}) is applied to the initial state of the system-environment
$\hat{\rho}(t_0)\otimes\hat{\rho}_{_E}(t_0)$ and then the tracing over
the environment is performed, so that $\hat{\rho}(t_k)$ can be written as
\be
\hat{\rho}(t_k)= \sum_{i_1,i_2=0}^N c_{i_1} (c_{i_2})^*
\hat{R}_{(i_1,i_2)}(t_k),
\label{5}
\ee
where	$(N+1)^2$ operators $\hat{R}_{(i_1,i_2)}(t_k)$ are defined
as
\be
\hat{R}_{(i_1,i_2)}(t_k)= \sum_{j_1,j_2=0}^N D_{(i_1,i_2)(j_1,j_2)}(t_k)
|j_1\rangle\langle j_2|,
\label{6}
\ee
with
\be
D_{(i_1,i_2)(j_1,j_2)}(t_k) & = &
\sum_{\alpha_1,\alpha_2,\gamma}
d_{\alpha_1\alpha_2}
\nonumber\\
 & \times &
E_{(i_1j_1)(\alpha_1\gamma)}(t_k)E^*_{(i_2j_2)(\alpha_2\gamma)}(t_k).
\label{7}
\ee
From Eq.(\ref{5}) it follows that the process $\hat{\cal T}(t_k)$
for a given time $t_k$ is {\it completely} determined by
$(N+1)^2$ operators $\hat{R}_{(i_1,i_2)}(t_k)$, which in turn are
specified by the $(N+1)^2\times (N+1)^2$ matrix elements
$D_{(i_1,i_2)(j_1,j_2)}(t_k)$.
We note that$\hat{R}_{(i_1,i_2)}(t_k)$	have the properties
\be
{\rm Tr} \hat{R}_{(i_1,i_2)}(t_k) & = & \delta_{i_1,i_2};\nonumber
\\
(\hat{R}_{(i_1,i_2)}(t_k))^{\dagger} & = & \hat{R}_{(i_2,i_1)}(t_k),
\label{8}
\ee
or, equivalently,
\be
\sum_{j=0}^N D_{(i_1,i_2)(j,j)}(t_k) & = & \delta_{i_1,i_2};\nonumber\\
D^*_{(i_1,i_2)(j_1,j_2)}(t_k) & = & D_{(i_2,i_1)(j_2,j_1)}(t_k).
\label{9}
\ee
We also note that neither $\hat{R}_{(i_1,i_2)}(t_k)$ nor
$D_{(i_1,i_2)(j_1,j_2)}(t_k)$
 depend
on the initial state $\hat{\rho}(t_0)$ of the system and formally
they fulfill the conditions
\be
\lim_{t_k\rightarrow t_0} \hat{R}_{(i_1,i_2)}(t_k) & = & |i_1\rangle\langle i_2|;
\nonumber \\
\lim_{t_k\rightarrow t_0}
D_{(i_1,i_2)(j_1,j_2)}(t_k) & = & \delta_{i_1,j_1}\delta_{i_2,j_2}.
\label{10}
\ee

Poyatos et al. \cite{Poyatos} have shown that
in order to specify the $(N+1)^2$ operators  $\hat{R}_{(i_1,i_2)}(t_k)$
one has to consider $(N+1)^2$ specific (see below) initial
conditions $|\Psi^{(k_1,k_2)}\rangle_{in}
=\sum_{i_1=0}^N c_{i_1}^{(k_1,k_2)}| i_1\rangle$ where
$k_1,k_2=0,1,...,N$
and to measure the corresponding $(N+1)^2$
output density operators $\hat{\rho}^{(k_1,k_2)}(t_k)$
which can be expressed as
\be
\hat{\rho}^{(k_1,k_2)}(t_k)= \sum_{i_1,i_2=0}^N
M_{(k_1,k_2)(i_1,i_2)}
\hat{R}_{(i_1,i_2)}(t_k),
\label{11}
\ee
where
\be
M_{(k_1,k_2)(i_1,i_2)}= c_{i_1}^{(k_1,k_2)} (c_{i_2}^{(k_1,k_2)})^* .
\label{12}
\ee
If the $(N+1)^2$ initial conditions $|\Psi^{(k_1,k_2)}\rangle_{in} $
are chosen so, that the matrix $M_{(k_1,k_2)(i_1,i_2)}$ given by
Eq. (\ref{12}) is invertible, then the set of Eqs.(\ref{11})
can be solved with respect of the operators
$\hat{R}_{(i_1,i_2)}(t_k)$. Alternatively, one can express the
matrix elements $D_{(i_1,i_2)(j_1,j_2)}(t_k)$ as functions of the
{\it in} and {\it out} states of the measured system, i.e.
\be
D_{(i_1,i_2)(j_1,j_2)}(t_k)
= \nonumber \\
 \sum_{k_1,k_2=0}^N \tilde{M}_{(i_1,i_2)(k_1,k_2)}
S_{(k_1,k_2)(j_1,j_2)}(t_k),
\label{13}
\ee
where the $(N+1)^2\times (N+1)^2$ matrix $S$ is defined as
\be
S_{(k_1,k_2)(j_1,j_2)}(t_k)= \langle j_1|
\hat{\rho}^{(k_1,k_2)}(t_k)|j_2\rangle.
\label{14}
\ee
The matrix $\tilde{M}$
is the inverse of $M$ and has the property
$
\sum_{k_1,k_2=0} \tilde{M}_{(j_1,j_2)(k_1,k_2)}
M_{(k_1,k_2)(i_1,i_2)}= \delta_{i_1,j_1} \delta_{i_2,j_2}.
$
So this is how the process $\hat{\cal T}(t_k)$ can be reconstructed from
the measured {\it in} and {\it out} states. To make the reconstruction
possible the matrix $M$ has to be invertible. Obviously, there are many
choices of such matrix. In particular, Poyatos et al. \cite{Poyatos}
have proposed $M$ given by Eq.(\ref{12})
with  complex amplitudes $c_{i}^{(k_1,k_2)}$
specified  as
\be
c_{i}^{(k_1,k_2)}=\left\{
\begin{array}{lcr}
 (\delta_{i,k_1} +\delta_{i,k_2})/\sqrt{2} & \mbox{\rm if} & k_1 > k_2 \\
\delta_{i,k_1}	& \mbox{\rm if} & k_1 = k_2 \\
(\delta_{i,k_1} + i \delta_{i,k_2})/\sqrt{2} & \mbox{\rm if} & k_1 < k_2
\end{array}
\right. .
\label{15}
\ee
The reconstruction process described above gives us  a set
of operators $\hat{R}_{(i_1,i_2)}(t_k)$ which describe the transition
of the system from the state $\hat{\rho}(t_0)$ to the state
$\hat{\rho}(t_k)$ at a given  time $t_k$.
In principle, one can perform a whole sequence of such reconstructions
at different times $t_1, t_2, .... t_K$ so that 
 the {\it reduced dynamics} of the
studied system can be reconstructed from the measured data.

Now our task is to determine
(reconstruct)
from a set of measurements of the output states $\hat{\rho}^{(k_1,k_2)}(t)$
for given input states	$\hat{\rho}^{(k_1,k_2)}(t_0)$,
the form of the Liouvillian superoperator
$\hat{\cal L}(t)$ in Eq.(\ref{3}). To do so, we firstly note, that 
when  the time evolution of the operators
 $\hat{\rho}^{(k_1,k_2)}(t)$ is governed by Eq.(\ref{3}), then taking into
account the expression (\ref{11}) and the assumption that the
matrix $M$ is invertible, we find that the operators
$\hat{R}_{(i_1,i_2)}(t)$ are also governed by the same master equation, i.e.
\be
\frac{d}{d t} \hat{R}_{(i_1,i_2)}(t)
= \hat{\cal L}(t) \hat{R}_{(i_1,i_2)}(t),
\label{16}
\ee
with the initial conditions given by Eq.(\ref{10}). Alternatively,
taking into account the expression (\ref{6}) we obtain from
Eq.(\ref{16}) a set of linear differential equations
for matrix elements $D_{(i_1,i_2)(k_1,k_2)}(t)$
\be
\frac{d}{d t} D_{(i_1,i_2)(k_1,k_2)}(t)
 =
\nonumber\\
 \sum_{j_1,j_2=0}^N D_{(i_1,i_2)(j_1,j_2)}(t)
G_{(j_1,j_2)(k_1,k_2)}(t),
\label{17}
\ee
with the initial conditions (\ref{10}). Here the matrix
$G_{(j_1,j_2)(k_1,k_2)}(t)$ is defined as
\be
G_{(j_1,j_2)(k_1,k_2)}(t)
=\langle k_1|\left( \hat{\cal L}(t) | j_1\rangle\langle j_2|\right)
| k_2\rangle,
\label{18}
\ee
and it {\it uniquely} determines the Liouvillian superoperator
$\hat{\cal L}(t)$.

We already know how to reconstruct  matrices $D$ from the
measured data for arbitrary time $t$ (from these data we can
 also evaluate the corresponding time derivatives).
Providing the matrix $D_{(i_1,i_2)(j_1,j_2)}(t)$ is not singular
 its inverse $\tilde{D}_{(j_1,j_2)(i_1,i_2)}(t)$ can be found and
then the reconstructed matrix $G_{(j_1,j_2)(k_1,k_2)}(t)$
is given by a simple expression
\be
G_{(j_1,j_2)(k_1,k_2)}(t)  =
\nonumber\\
 \sum_{i_1,i_2=0}^N \tilde{D}_{(j_1,j_2)(i_1,i_2)}(t)\frac{d}{d t}
 D_{(i_1,i_2)(k_1,k_2)}(t)
\label{19}
\ee
from which the superoperator $\hat{\cal L}(t)$
at time $t$ can be determined. This is the main result of the paper.

In the following we will apply this general algorithm into three physically
interesting examples.

{\it Example A.} --
Let us consider a two-level system (a two-level atom, a spin-1/2,
or a qubit) with a two-dimensional Hilbert space
${\cal H}_{_S}$ spanned by two vectors $|1\rangle$ and $|0\rangle$.
In order to specify the Liouvillian superoperator $\hat{\cal L}(t)$
for the two-level atom 
we have to know the time evolution of four
initial states  specified by Eq.(\ref{15}).
Let us assume that from the measured data it is found that these states
evolve as
\be
\hat{\rho}^{(0,0)}(t) & = &
\left(
\begin{array}{cc}
0 & 0 \\
0 & 1
\end{array}
\right);    \qquad
\hat{\rho}^{(1,1)}(t)  =
\left(
\begin{array}{cc}
e^{-\Gamma t} & 0 \\
0 & 1-e^{-\Gamma t}
\end{array}
\right).
\nonumber\\
\hat{\rho}^{(0,1)}(t) & = & \frac{1}{2}
\left(
\begin{array}{cc}
e^{-\Gamma t} & ie^{-\Gamma t/2} \\
-ie^{-\Gamma t/2} & 2- e^{-\Gamma t}
\end{array}
\right);
\nonumber\\
\hat{\rho}^{(1,0)}(t) & = & \frac{1}{2}
\left(
\begin{array}{cc}
e^{-\Gamma t} & e^{-\Gamma t/2} \\
e^{-\Gamma t/2} & 2-e^{-\Gamma t}
\end{array}
\right).
\label{22}
\ee
Now we can apply our reconstruction scheme and we find for the
matrix $G_{(j_1,j_2)(k_1,k_2)}(t)$ the expression \cite{comment}
\be
G_{(j_1,j_2)(k_1,k_2)}(t)=\left(
\begin{array}{cccc}
-\Gamma  & 0 & 0 & \Gamma  \\
0 &  -\Gamma /2
& 0 & 0 \\
0 & 0
&  -\Gamma /2 & 0\\
0 & 0 & 0 & 0
\end{array}
\right).
\label{23}
\ee
This matrix  corresponds to the
Liouvillian which defines the  master equation
\be
\frac{d}{d t} \hat{\rho}=
\hat{\cal L} \hat{\rho} = \frac{\Gamma}{2}\left[2 \hat{\sigma}_-
\hat{\rho}\hat{\sigma}_+ - \hat{\sigma}_+\hat{\sigma}_-\hat{\rho}
- \hat{\rho} \hat{\sigma}_+\hat{\sigma}_-\right],
\label{24}
\ee
describing the decay of a two-level atom into a zero-temperature
reservoir \cite{Louisell}. The Liouvillian in Eq.(\ref{24}) 
is time independent which reflects the fact that the state of
the reservoir does not change  in time under the influence of
the system.

{\it Example B.} --
Here we will reconstruct the Liouvillian superoperator for the master
equation describing the time evolution of a single two-level atom
interacting with a single-mode electro-magnetic field in an ideal
cavity. The corresponding  Hamiltonian in the dipole and the rotating-wave
approximations  reads \cite{Louisell}
\be
\hat{H}= \omega_{_A}\hat{\sigma}_z +\omega\hat{a}^{\dagger}\hat{a}+
\lambda (\hat{\sigma}_+\hat{a}+\hat{\sigma}_-\hat{a}^{\dagger}),
\label{25}
\ee
where $\lambda$ is the atom-field coupling constant. 
We assume that the atomic transition frequency ($\omega_{_A}$)
 is on the resonance with the filed frequency ($\omega$).
 The operators $\hat{a}^{\dagger}$
and $\hat{a}$ are the usual photon creation and annihilation
operators, respectively, with $[\hat{a},\hat{a}^{\dagger}]=1$.
If the atom  and the field are initially
prepared  in states
$
|\Psi(t_0)\rangle_{_A}  =  c_0 |0\rangle + c_1 | 1\rangle,
$ and $
|\Psi(t_0)\rangle_{_F}  =  \sum_{k=0}
e_k | k\rangle \equiv |\alpha\rangle,
$
respectively, then at time $t$ the atom-field state vector
$|\Psi(t)\rangle_{_{A-F}}$ reads
\be
|\Psi(t)\rangle_{_{A-F}}  =
c_{0}\sum_{k}\left(\cos\tau_k |k\rangle |0\rangle 
- i\sin\tau_k |k-1\rangle|1\rangle\right)
\nonumber \\ 
+  c_{1}\sum_{k}\left(\cos\tau_{k+1} |k\rangle |1\rangle 
- i\sin\tau_{k+1} |k+1\rangle |0\rangle\right),
\label{27}
\ee
where $\tau_k=\lambda\sqrt{k} t$.
Utilizing Eq.(\ref{19})
we can determine  the Liouvillian superoperator which governs
the dynamics of the atom. Here  $\hat{\cal L}(t)$
explicitly depends on the initial state of the cavity field.
Let us assume a particular case when the field has been
prepared in the Fock state $|M\rangle$. With this initial state
the matrix (\ref{13}) takes the form
\be
D_{(i_1,i_2)(j_1,j_2)}(t)=\left(
\begin{array}{cccc}
\xi_1  & 0 & 0 & 1- \xi_1  \\
0 &  \sqrt{\xi_0\xi_1}
& 0 & 0 \\
0 & 0
&  \sqrt{\xi_0\xi_1} & 0\\
1- \xi_0 & 0 & 0 & \xi_0
\end{array}
\right),
\label{30}
\ee
where $\xi_0= \cos^2(\lambda t \sqrt{M})$ and
$\xi_1 = \cos^2(\lambda t \sqrt{M+1})$.
The determinant of this matrix $det[D]=\xi_0\xi_1(\xi_0+\xi_1-1)$
is equal to zero only
at discrete moments so $D$ is invertible and
 we can utilize Eq.(\ref{19}) from which we find
\be
G_{(j_1,j_2)(k_1,k_2)}(t)
=\left(
\begin{array}{cccc}
-\gamma_1  & 0 & 0 & \gamma_1  \\
0 &  -\gamma_2/2
& 0 & 0 \\
0 & 0
&  -\gamma_2/2 & 0\\
\gamma_3 & 0 & 0 & -\gamma_3
\end{array}
\right),
\label{32}
\ee
with the time-dependent parameters $\gamma_i(t)$ given as
\end{multicols}
\vspace{-0.5cm}
\noindent\rule{0.5\textwidth}{0.4pt}\rule{0.4pt}{\baselineskip}
\widetext
\be
\gamma_1(t) & = & \frac{
2\lambda\left(\sqrt{M}\sin(2\lambda t\sqrt{M}) \sin^2(\lambda t \sqrt{M+1})
+\sqrt{M+1}\sin(2\lambda t \sqrt{M+1}) \cos^2(\lambda t \sqrt{M})
\right)}{[\cos(2\lambda t \sqrt{M})+\cos(2\lambda t \sqrt{M+1})]}
\nonumber\\
\gamma_2(t) & = & \frac{
\lambda\left(\sqrt{M}\sin(2\lambda t\sqrt{M}) \cos^2(\lambda t \sqrt{M+1})
+\sqrt{M+1}\sin(2\lambda t \sqrt{M+1}) \cos^2(\lambda t \sqrt{M})
\right)}{\cos^2(\lambda t \sqrt{M})\cos^2(\lambda t \sqrt{M+1})}
\nonumber\\
\gamma_3(t) & = & \frac{
2\lambda\left(\sqrt{M}\sin(2\lambda t\sqrt{M}) \cos^2(\lambda t \sqrt{M+1})
+\sqrt{M+1}\sin(2\lambda t \sqrt{M+1}) \sin^2(\lambda t \sqrt{M})
\right)}{[\cos(2\lambda t \sqrt{M})+\cos(2\lambda t \sqrt{M+1})]}.
\label{33}
\ee
\begin{multicols}{2}
From the solution (\ref{32})
 it follows that the Liouvillian superoperator is explicitly
time-dependent which reflects dynamical response 
of the environment
(i.e. the cavity field).
The master equation (\ref{3}) with $\hat{\cal L}(t)$ specified by
Eq.(\ref{32}) can be written as
\be
\frac{d}{d t}\hat{\rho} & = &
\frac{\gamma_1(t)}{2}\left[
2 \hat{\sigma}_-\hat{\rho}\hat{\sigma}_+
-\hat{\sigma}_+\hat{\sigma}_-\hat{\rho}
- \hat{\rho} \hat{\sigma}_+\hat{\sigma}_- \right]
\nonumber\\
 & - &
 \frac{\eta(t)}{2}\left[
\hat{\sigma}_+\hat{\sigma}_-\hat{\rho}\,\hat{\sigma}_-\hat{\sigma}_+
+
\hat{\sigma}_-\hat{\sigma}_+\hat{\rho}\,\hat{\sigma}_+\hat{\sigma}_-
\right]
\nonumber\\
 & + &
\frac{\gamma_3(t)}{2}\left[
 2\hat{\sigma}_+\hat{\rho}\hat{\sigma}_-
- \hat{\sigma}_-\hat{\sigma}_+\hat{\rho}
- \hat{\rho} \hat{\sigma}_-\hat{\sigma}_+ \right],
\label{34}
\ee
with the coefficients $\gamma_i(t)$
Eq.(\ref{33}) and
$\eta(t)=\gamma_2(t) - \gamma_1(t) -\gamma_3(t)$.
One can check that $\hat{\rho}_A(t)$ obtained from Eq.(\ref{27})
is the solution of the master equation (\ref{34}).
We note that if the cavity field is initially in the vacuum state
($M=0$) then the master equation (\ref{34}) takes the form (\ref{24})
but with the time-dependent ``decay'' rate $\Gamma\rightarrow
\gamma_1(t)=2\lambda
\tan \lambda t$.

\begin{figure}[]
\leavevmode
\centerline{\epsfig{height=5.0cm,file=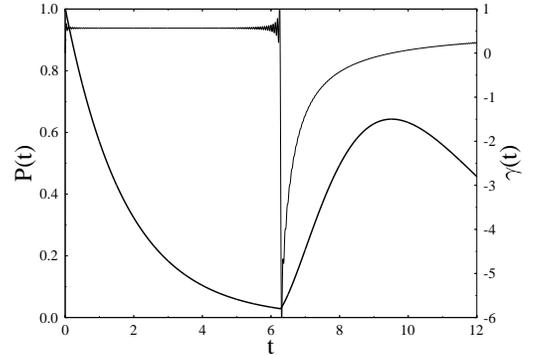}}
\caption{\narrowtext 
The time evolution of the decay rate $\gamma(t)$ (thin line)
and the  population of the excited
atomic level $P(t)$ (thick line).
We assume the atom to be in the center of the 1-D cavity, so it is coupled
only to the odd modes (i.e. $\lambda_{2k}=0$).
We assume $L=2\pi$ and $c=1$ so that
$\omega_{2k+1}=k+1/2$, and $\lambda_{2k+1}=\lambda=0.3$. The effective density
of modes which interact with the atom is $d_{eff}(\omega)=L/2c\pi=1$.
Therefore the decay rate 
$\Gamma=2\pi\lambda^2 d_{eff}(\omega)\simeq 0.564$. 
We consider 
 $K=400$ modes of the field  
 initially in the vacuum state and   
the atom (with $\omega_{_A}=101$)  in its upper state $|1\rangle$.
}
\label{fig1}
\end{figure}

{\it Example C.} --
Finally, we consider  a single two-level atom 
coupled to $K$ modes of the electro-magnetic field in a one-dimensional
cavity of the length $L$. The spectrum of modes is discrete with
frequencies $\omega_k = k \pi c/L$.
The corresponding
total Hamiltonian in the dipole and rotating-wave approximations reads
\cite{Cohen}
\be
\hat{H}= \omega_{_A}\hat{\sigma}_z +
\sum_{k=1}^K \omega_k\hat{a}^{\dagger}_k\hat{a}_k +  
\sum_{k=1}^K
\lambda_k (\hat{\sigma}_+\hat{a}_k+\hat{\sigma}_-\hat{a}^{\dagger}_k).
\label{35}
\ee
The field is assumed to be initially in
the {\it vacuum} state.  
By applying our algorithm we find the
master equation for the atom 
to be of the form (\ref{24}) except the "decay" rate
$\Gamma\rightarrow\gamma(t)$ 
is now explicitly time dependent. It can be expressed in terms
of the ``measured''
probability $P(t)=\langle 1|\hat{\rho}_{_A}(t)| 1\rangle$  
 that the upper atomic level is excited: 
\be
\gamma(t) = - \left(\frac{d P(t)}{d t}\right) P(t)^{-1}.
\label{36}
\ee
In Fig.\ref{fig1} we present 
the time evolution of $P(t)$ and $\gamma(t)$ obtained with the help of  
numerical diagonalization of the Hamiltonian (\ref{35}). From our results
it follows that $\gamma(t=0)=0$ but as soon as the atom starts to radiate
the function $\gamma(t)$ starts to grow and after a short time it takes
the constant value $\Gamma=2\pi\lambda^2 d_{eff}(\omega)$ given by the
Golden Fermi rule \cite{Cohen}. At this stage the atom radiates
exponentially and two wave packets propagating to the left and the right
cavity mirrors are irradiated. These packets are reflected by mirrors
at $t=L/2c$ and they ``kick'' back the atom at $t=L/c$. At this point
the atom is essentially in its ground state and the reflected waves packets
(environment) force it to absorb energy, i.e. the atom does not decay
exponentially anymore.
This is the reason why 
during the recurrence of the atomic inversion  $\gamma(t)$
rapidly changes and takes negative values.  

I thank Pavel B\'ona, Ignacio Cirac, Gabriel Drobn\'y, Jason Twamley, 
and Peter Zoller for helpful discussions
and comments. This work was supported by the {\it Aktion
\"{O}sterreich -- Slowakische Republik} under the project 18s42, and
by the Royal Society.


\end{multicols}


\begin{thebibliography}{99}
\bibitem{comment}
Matrix elements $G_{(j_1,j_2)(k_1,k_2)}$
are label so, that the pair of  $(j_1,j_2)$ indicate
lines [ the $j$-th line is numbered as
$j=(N+1)^2-(N+1)j_1-j_2$]
 while $(k_1,k_2)$ indicate rows [with $k$-th row numbered
as $k=(N+1)^2-(N+1)k_1-k_2$].
\end{thebibliography}
\end{document}